# Effective Analysis of C Programs by Rewriting Variability

Alexandru F. Iosif-Lazar[a], Jean Melo[a], Aleksandar S. Dimovski[a], Claus Brabrand[a], and Andrzej Wąsowski[a]

a    IT University of Copenhagen, Denmark

**Abstract**    Context. Variability-intensive programs (program families) appear in many application areas and for many reasons today. Different family members, called variants, are derived by switching statically configurable options (features) on and off, while reuse of the common code is maximized.
Inquiry. Verification of program families is challenging since the number of variants is exponential in the number of features. Existing single-program analysis and verification tools cannot be applied directly to program families, and designing and implementing the corresponding variability-aware versions is tedious and laborious.
Approach. In this work, we propose a range of variability-related transformations for translating program families into single programs by replacing compile-time variability with run-time variability (non-determinism). The obtained transformed programs can be subsequently analyzed using the conventional off-the-shelf single-program analysis tools such as type checkers, symbolic executors, model checkers, and static analyzers.
Knowledge. Our variability-related transformations are *outcome-preserving*, which means that the relation between the outcomes in the transformed single program and the union of outcomes of all variants derived from the original program family is *equality*.
Grounding. We present our transformation rules and their correctness with respect to a minimal core imperative language IMP. Then, we discuss our experience of implementing and using the transformations for efficient and effective analysis and verification of real-world C program families.
Importance. We report some interesting variability-related bugs that we discovered using various state-of-the-art single-program C verification tools, such as Frama-C, Clang, LLBMC.

# The Art, Science, and Engineering of Programming

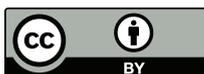

# Effective Analysis of C Programs by Rewriting Variability

## 1 Introduction

Many software systems today are variability intensive. They permit users to derive a custom variant by choosing suitable configuration options (features) depending on their requirements. There are different strategies for implementing variational systems (program families) [11]. Still, many popular industrial program families from system software (e.g. Linux kernel) and embedded software (e.g. cars, phones, avionics) domains are implemented using annotative approaches such as conditional compilation. For example, `#ifdef` annotations from the C-preprocessor are used to specify under which conditions, parts of the code should be included or excluded from a variant.

Due to the increasing popularity of program families, formal verification techniques for proving their correctness are widely studied (see [35] for a survey). Analyzing program families is challenging [29]. From only a few compile-time configuration options, exponentially many variants can be derived. Thus, for large variability-intensive software systems, any brute-force approach that derives and analyzes all variants individually one by one using existing single-program analysis tools is very inefficient or even infeasible. Recently, many dedicated family-based (variability-aware) analysis tools have been developed, which operate directly on program families. They produce results for all variants at once in a single run by exploiting the similarities between the variants. Examples of successful family-based analysis tools are applied to syntax checking [25, 20], type checking [24, 8], static analysis [7, 6], model checking [10, 14], etc. Although they are more efficient than the brute-force approach, still their design and implementation for each particular analysis and language is tedious and error prone. Often, these family-based tools are research prototypes implemented from scratch. So it is very difficult to re-implement all optimization algorithms in them that already exist for their single-program industrial-strength counterparts, which have been under development for several decades.

Another approach for efficient variability-aware verification would be to replace compile-time variability with run-time variability (or non-determinism) [37]. In particular, in this work we consider a class of variability-related transformations that transform a program family into a single program, whose outcomes are equal to the union of all outcomes of individual variants. We call the corresponding transformations outcome-preserving. Subsequently, existing single-program analysis tools (verification oracles) that can handle non-determinism (run-time variability) can be used to analyze the generated single program. Finally, the obtained results are interpreted back on the individual variants. The overview of this approach is given in Figure 1. Instead of using specialized variability-aware tools to analyze program families (which would be tedious and labor intensive), our transformation-based approach allows us to use the standard off-the-shelf single-program analysis tools to achieve the same goal. Nevertheless, the limitation of our approach is that we may not obtain the most precise conclusive results for all individual variants. Of course, this depends on the particular analysis and tool that we use.

To demonstrate correctness of our transformation-based approach, we define the transformations formally using IMP, a small imperative language. To model compile-time variability, we extend IMP with an "#ifdef" construct for encoding multiple variants, which we call $\overline{\text{IMP}}$ language. To encode run-time variability, we extend IMP with an "or" construct for encoding non-determinism, which we call IMPor language. We define transformations that translate any given $\overline{\text{IMP}}$ program into a corresponding IMPor program. Furthermore, for each transformation we show the relation between the semantics of the input and output programs.

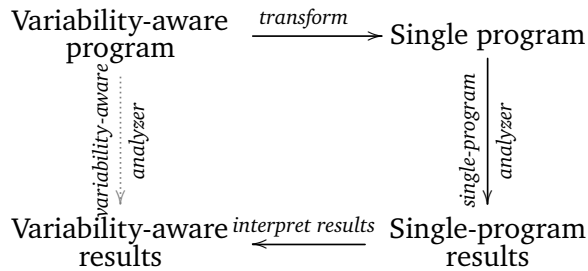

**Figure 1** The overview of our transformation-based approach for verification of program families. The single-program analyzer can be any verification oracle for single programs, such as: symbolic executor, type checker, static analyzer, model checker.

Finally, we report on our experience with implementing and applying our transformations for a full-fledged language, C. The tool, called C Reconfigurator, uses variability-aware parser SuperC [20] for parsing C code with preprocessor annotations, then applies our variability rewrites thus producing a single C program as output. We evaluate our approach on real-world variability intensive C programs with real bugs. We show how some known off-the-shelf single-program analysis tools can be used for efficient and effective verification of such programs.

In summary, this work makes the following contributions:

- A stand-alone variability-related transformation, which transforms a program family into a single program by replacing compile-time variability with non-determinism.
- Correctness of the proposed transformation, which shows that the set of outcomes of the transformed program is equal to the union of sets of outcomes of variants from the input family.
- A prototype tool, C Reconfigurator, which implements the above variability-related transformation for the C language.
- An evaluation of the effectiveness of our transformation-based approach for finding real variability bugs in large variability intensive C software systems.

## 2 Motivating Example

We begin by showing how our variability transformations work on C program families. Consider a preprocessor-based family of C programs shown in Figure 2 (left column), which uses two (Boolean) features $A$ and $B$. Our two features give rise to a family of four variants defined by the set of configurations $\mathbb{K} = \{A \wedge B, A \wedge \neg B, \neg A \wedge B, \neg A \wedge \neg B\}$.

**Effective Analysis of C Programs by Rewriting Variability**

| | |
|---|---|
| int foo() { | int $A := rand() \% 2$; |
|    int x:= 1; | int $B := rand() \% 2$; |
|    #if (A) x:= x+1#endif; | int foo() { |
|    #if (B) x:= x-1#endif; |    int x := 1; |
|    return 2/x; |    if (A) x:= x+1; |
| } |    if (B) x:= x-1; |
| |    return 2/x; } |

■ **Figure 2** Before (left column) and after (right column) our transformations

For each configuration a different variant (single program) can be generated by appropriately resolving #if statements. For example, the variant for $A \wedge B$ will have both features $A$ and $B$ enabled (set to true), thus yielding the following body of $foo()$: int x := 1; x := x+1; x := x-1; return 2/x. The variant for $\neg A \wedge \neg B$ is: int x := 1; return 2/x. In such program families, errors (also known as *variability bugs* [1]) can occur in some variants (configurations) but not in others. In our example program family in Figure 2, the variant $\neg A \wedge B$ will crash at the return statement when we attempt to divide by zero. On the other hand, the other variants do not contain the division-by-zero error since the value of x at the return statement is: 1 for variants $A \wedge B$ and $\neg A \wedge \neg B$, and 2 for $A \wedge \neg B$.

In Figure 2, we show a single program (right column) obtained by applying our variability-related transformation on the family shown in the left column. All features are first declared as ordinary global variables and non-deterministically initialized to 0 or 1, then all #if statements are transformed into ordinary if-s with the same conditions. Thus, the division-by-zero error is present in this single program and corresponds to a trace when $A$ is initialized to 0 and $B$ to 1. The set of outcomes of the transformed program (Figure 2, right column) is equal to the union of outcomes of all individual variants from the family (Figure 2, left column). Therefore, the division-by-zero error is present in the transformed program.

In general, the transformed program that we obtain from the original program family can be analyzed by various single-program verification tools, in order to find variability errors or to confirm the absence of errors in the given program family.

## 3 A Formal Model for Transformations

We now introduce the IMP language that we use to demonstrate our transformations and their proofs of correctness. We describe two extensions of IMP: $\underline{\text{IMP}}$or used to represent run-time variability (non-determinism), and $\overline{\text{IMP}}$ used to represent compile-time variability.

A. F. Iosif-Lazar, J. Melo, A. S. Dimovski, C. Brabrand, A. Wasowski### 3.1 IMP

We use a simple imperative language, called IMP [32, 34], which represents a regular general-purpose programming language, aimed at the development of single programs. IMP is a well established minimal language, which is used only for presentational purposes here.

**Syntax.** IMP is an imperative language with two syntactic categories: expressions and statements. Expressions include integer constants, variables, and binary operations. Statements include a "do-nothing" statement skip, assignments, statement sequences, conditional statements, while loops, and local variable declarations. Its abstract syntax is summarized using the following grammar:

$$e \;::=\; n \;|\; \mathrm{x} \;|\; e_0 \oplus e_1$$
$$s \;::=\; \mathrm{skip}\,|\,\mathrm{x} := e\,|\,s_0\,;\,s_1\,|\,\mathrm{if}\ e\ \mathrm{then}\ s_0\ \mathrm{else}\ s_1\,|\,\mathrm{while}\ e\ \mathrm{do}\ s\,|\,\mathrm{var}\ \mathrm{x}{:=}e\ \mathrm{in}\ s$$

In the above, $n$ stands for an integer constant, x stands for a variable name, and $\oplus$ stands for any binary arithmetic operator. We denote by *Stm* and *Exp* the set of all statements, $s$, and expressions, $e$, generated by the above grammar.

**Semantics.** A state of a program is a *store* mapping variables to values (integer numbers), $Val = \mathbb{Z}$. We write $Store = Var \to Val$ to denote the set of all possible stores. Expressions are computed in a given store, denoted by $\sigma$. A function $\mathcal{E} : Exp \times Store \to Val$ defined below by induction on $e$, maps an expression and a store to a single value, thereby formalizing evaluation of expressions.

$$\mathcal{E}(n,\sigma) = n, \qquad \mathcal{E}(\mathrm{x},\sigma) = \sigma(\mathrm{x}), \qquad \mathcal{E}(e_0 \oplus e_1,\sigma) = \mathcal{E}(e_0,\sigma) \oplus \mathcal{E}(e_1,\sigma)$$

Figure 3 presents the inference rules for a small-step operational semantics for IMP [32, 34]. The notation $\sigma[\mathrm{x} \mapsto n]$ denotes the state that maps x into $n$ and all other variables y into $\sigma(\mathrm{y})$. Following the convention popularized by C, we model Boolean values as integers, with zero interpreted as false and everything else as true (see rules If2 and Wh2, respectively, If1 and Wh1). Note that for variable declarations (see rules Var1 or Var2) we need to restore the declared variable, x, to its earlier global value assigned to x before the declaration, when the scope of declaration has completed. That is why the statement $s'$ in intermediate configurations (rule Var1) is prefixed with variable declarations whose initializations store the local values of x. We can use the inference rules in Figure 3 to define the transition relation: $\langle s, \sigma \rangle \to \gamma$, where $\gamma$ is either of the form $\langle s', \sigma' \rangle$ or of the form $\sigma'$. If $\gamma$ is of the form $\langle s', \sigma' \rangle$ then the execution of $s$ is not completed and the complex statement $s$ is rewritten into simpler one $s'$, possibly updating the store $\sigma$ into $\sigma'$ (for instance, Seq1 or Seq2). If $\gamma$ is of the form $\sigma'$ then the execution of $s$ from $\sigma$ has terminated and the final state is $\sigma'$ (for instance, Skip or Wh2).

A *derivation sequence* of $s$ starting in store $\sigma$ can be either a finite sequence $\langle s, \sigma \rangle \to \langle s_1, \sigma_1 \rangle \to \ldots \to \sigma'$ (means: $s$ is run in $\sigma$ and terminates successfully transforming $\sigma$ to $\sigma'$ in the process), or an infinite sequence $\langle s, \sigma \rangle \to \langle s_1, \sigma_1 \rangle \to \ldots$ (means: $s$ diverges

$$\text{Skip} \frac{}{\langle \text{skip}, \sigma \rangle \to \sigma} \quad \text{Asgn} \frac{n = \mathcal{E}(e, \sigma)}{\langle x := e, \sigma \rangle \to \sigma[x \mapsto n]} \quad \text{Sq1} \frac{\langle s_0, \sigma \rangle \to \langle s_0', \sigma' \rangle}{\langle s_0; s_1, \sigma \rangle \to \langle s_0'; s_1, \sigma' \rangle}$$

$$\text{Sq2} \frac{\langle s_0, \sigma \rangle \to \sigma'}{\langle s_0; s_1, \sigma \rangle \to \langle s_1, \sigma' \rangle} \quad \text{If1} \frac{\mathcal{E}(e, \sigma) \neq 0}{\langle \text{if } e \text{ then } s_0 \text{ else } s_1, \sigma \rangle \to \langle s_0, \sigma \rangle}$$

$$\text{If2} \frac{\mathcal{E}(e, \sigma) = 0}{\langle \text{if } e \text{ then } s_0 \text{ else } s_1, \sigma \rangle \to \langle s_1, \sigma \rangle} \quad \text{Wh1} \frac{\mathcal{E}(e, \sigma) \neq 0}{\langle \text{while } e \text{ do } s, \sigma \rangle \to \langle s; \text{while } e \text{ do } s, \sigma \rangle}$$

$$\text{Wh2} \frac{\mathcal{E}(e, \sigma) = 0}{\langle \text{while } e \text{ do } s, \sigma \rangle \to \sigma} \quad \text{Var1} \frac{n = \mathcal{E}(e, \sigma) \quad \langle s, \sigma[x \mapsto n] \rangle \to \langle s', \sigma' \rangle}{\langle \text{var } x := e \text{ in } s, \sigma \rangle \to \langle \text{var } x := \sigma'(x) \text{ in } s', \sigma'[x \mapsto \sigma(x)] \rangle}$$

$$\text{Var2} \frac{n = \mathcal{E}(e, \sigma) \quad \langle s, \sigma[x \mapsto n] \rangle \to \sigma'}{\langle \text{var } x := e \text{ in } s, \sigma \rangle \to \sigma'[x \mapsto \sigma(x)]}$$

**Figure 3** Small-step operational semantics for IMP

when run in $\sigma$). We write $[\![s]\!]\sigma$ for the final store $\sigma'$ that can be derived from $\langle s, \sigma \rangle$ (if the derivation is finite), i.e. $\langle s, \sigma \rangle \to^* \sigma'$, otherwise if the derivation is infinite $[\![s]\!]\sigma$ is undefined (empty). In general, we define:

$$[\![s]\!] = \bigcup_{\sigma \in \text{Store}^{\text{Init}}} [\![s]\!]\sigma$$

where $\text{Store}^{\text{Init}}$ denotes the set of initial input stores on which $s$ is executed.

## 3.2 IMPor

**Syntax** The language IMPor is obtained by extending IMP with a non-deterministic choice operator 'or' which can non-deterministically choose to evaluate either of its arguments.

$$e \quad ::= \quad \ldots \mid e_0 \text{ or } e_1$$

**Semantics.** Since we have non-deterministic construct 'or', it is possible for an expression to evaluate to a set of different values in a given store. Therefore, now we have $\mathcal{E} : \text{Exp} \times \text{Store} \to \mathcal{P}(\text{Val})$ defined as follows:

$$\mathcal{E}(n, \sigma) = \{n\}, \qquad \mathcal{E}(x, \sigma) = \{\sigma(x)\}, \qquad \mathcal{E}(e_0 \text{ or } e_1, \sigma) = \mathcal{E}(e_0, \sigma) \cup \mathcal{E}(e_1, \sigma)$$
$$\mathcal{E}(e_0 \oplus e_1, \sigma) = \{v_0 \oplus v_1 \mid v_0 \in \mathcal{E}(e_0, \sigma), v_1 \in \mathcal{E}(e_1, \sigma)\}$$

The small-step operational semantics rules for IMPor are those of IMP given in Figure 3, but now they take into account the non-determinism of $\mathcal{E}(e, \sigma)$. For example, we have:

$$\text{Wh1} \frac{n \in \mathcal{E}(e, \sigma) \quad n \neq 0}{\langle \text{while } e \text{ do } s, \sigma \rangle \to \langle s; \text{while } e \text{ do } s, \sigma \rangle} \quad \text{Wh2} \frac{0 \in \mathcal{E}(e, \sigma)}{\langle \text{while } e \text{ do } s, \sigma \rangle \to \sigma}$$

For IMPor, we write $[\![s]\!]\sigma$ for the *set* of final stores $\sigma'$ that can be derived from $\langle s, \sigma \rangle$, i.e. $\langle s, \sigma \rangle \to^* \sigma'$. Note that since IMPor is a non-deterministic language $[\![s]\!]\sigma$ may contain more than one final store. Finally, $[\![s]\!] = \bigcup_{\sigma \in \text{Store}^{\text{Init}}} [\![s]\!]\sigma$.

## 3.3 $\overline{\text{IMP}}$

A finite set of Boolean variables $\mathbb{F} = \{A_1, \ldots, A_n\}$ describes the set of available *features* in the program family. Each feature may be *enabled* or *disabled* in a particular variant.

A. F. Iosif-Lazar, J. Melo, A. S. Dimovski, C. Brabrand, A. Wasowski

A *configuration* $k$ is a truth assignment or a valuation which gives a truth value to each feature, i.e. $k$ is a mapping from $\mathbb{F}$ to {true, false}. If a feature $A \in \mathbb{F}$ is enabled for the configuration $k$ then $k(A) = \text{true}$, otherwise $k(A) = \text{false}$. Any configuration $k$ can also be encoded as a conjunction of literals: $k(A_1) \cdot A_1 \wedge \cdots \wedge k(A_n) \cdot A_n$, where $\text{true} \cdot A = A$ and $\text{false} \cdot A = \neg A$. We write $\mathbb{K}$ for the set of all *valid* configurations defined over $\mathbb{F}$ for a family. The set of valid configurations is typically described by a feature model [23], but in this work we disregard syntactic representations of the set $\mathbb{K}$. Note that $|\mathbb{K}| \leq 2^{|\mathbb{F}|}$, since, in general, not every combination of features yields a *valid* configuration. We define *feature expressions*, denoted *FeatExp*, as the set of well-formed propositional logic formulas over $\mathbb{F}$ generated using the grammar:
$\phi ::= \text{true} \,|\, A \in \mathbb{F} \,|\, \neg \phi \,|\, \phi_1 \wedge \phi_2 \,|\, \phi_1 \vee \phi_2$.

**Syntax.** The programming language $\overline{\text{IMP}}$ is our two-stage extension of IMP (thus, $\overline{\text{IMP}}$ does not contain the 'or' construct). Its abstract syntax includes the same expression and statement productions as IMP, but we add the new compile-time conditional statements for encoding multiple variants of a program. The new statements "#if ($\phi$) $s$ #endif" and "#if ($\phi$) var x:=$n$ in #endif $s$" contain a feature expression $\phi \in \textit{FeatExp}$ as a presence condition, such that only if $\phi$ is satisfied by a configuration $k \in \mathbb{K}$ then the code between #if and #endif will be included in the variant for $k$.

$s ::= \ldots \,|\, \text{\#if}\,(\phi)\,s\,\text{\#endif} \,|\, \text{\#if}\,(\phi)\,\text{var x:=}n\,\text{in \#endif}\,s$

Note that only statements and local variable declarations can be compile-time conditionally defined in $\overline{\text{IMP}}$. However, in general "#if" constructs defined on arbitrary language elements could be translated into constructs that respect the appropriate syntactic structure of the language by code duplication [19]. Also note that the C preprocessor uses the following keywords: #if, #ifdef, and #ifndef to start a conditional construct; #elif and #else to create additional branches; and #endif to end a construct. Any of such preprocessor conditional constructs can be desugared and represented only by #if construct we use in this work, e.g. #ifdef ($\phi$) $s_0$ #else $s_1$ #endif is translated into #if ($\phi$) $s_0$ #endif ; #if ($\neg\phi$) $s_1$ #endif.

**Semantics.** The semantics of $\overline{\text{IMP}}$ has two stages: first, given a configuration $k \in \mathbb{K}$ compute an IMP single program without #if-s; second, evaluate the obtained variant using the standard IMP semantics. The first stage is a simple *preprocessor* specified by the projection function $\pi_k$ mapping an $\overline{\text{IMP}}$ program family into an IMP single program corresponding to the configuration $k \in \mathbb{K}$. The projection $\pi_k$ copies all basic statements of $\overline{\text{IMP}}$ that are also in IMP, and recursively pre-processes all sub-statements of compound statements. For example, $\pi_k(\text{skip}) = \text{skip}$ and $\pi_k(s_0;s_1) = \pi_k(s_0);\pi_k(s_1)$. The interesting case is "#if ($\phi$) $s$ #endif" (resp., #if ($\phi$) var x:=$n$ in #endif $s$) statement, where the statement $s$ (resp., the local variable declaration var x:=n in) is included in the resulting variant iff $k \models \phi$ [1], otherwise it is removed. We have:

$$\pi_k(\text{\#if } (\phi) \ s \ \text{\#endif}) = \begin{cases} \pi_k(s) & \text{if } k \models \phi \\ \text{skip} & \text{if } k \not\models \phi \end{cases}$$

$$\pi_k(\text{\#if } (\phi) \text{ var x:=n in \#endif } s) = \begin{cases} \pi_k(\text{var x:=n in } s) & \text{if } k \models \phi \\ \pi_k(s) & \text{if } k \not\models \phi \end{cases}$$

Note that since any configuration $k \in \mathbb{K}$ has only one satisfying truth assignment (values of all features are fixed in $k$), either $k \models \phi$ or $k \not\models \phi$ for any $\phi \in \textit{FeatExp}$.

## 4   Variability-related Transformations

Our aim is to transform an input $\overline{\text{IMP}}$ program family $\overline{s}$ with sets of features $\mathbb{F}$ and configurations $\mathbb{K}$ into an output IMPor program $\overline{s'}$.

In a pre-transformation phase, we first convert each feature $A \in \mathbb{F}$ into the variable $A$, which is non-deterministically initialized to 0 or 1 (meaning to false or true). Let $\mathbb{F} = \{A_1, \ldots, A_n\}$ be the set of available features in the family $\overline{s}$, then we have the following initialization fragment in the resulting pre-transformed program pre-t($\overline{s}$):

   pre-t($\overline{s}$) = var $A_1$:=0 or 1 in ... var $A_n$:=0 or 1 in $\overline{s}$

Note that in the initialization we consider all possible combination of values for features (in total $2^{|\mathbb{F}|}$). We will take into account the specific set of configurations $\mathbb{K}$ ($|\mathbb{K}| \leq 2^{|\mathbb{F}|}$) later on, in the transformation phase.

In the following, rewrite rules have the form:

$\psi \vdash s \leadsto s'$

meaning that: if the current program family being transformed matches any abstract syntax tree (AST) node of the shape $s$ nested under #if-s with the resulting presence condition that implies $\psi \in \textit{FeatExp}$ (i.e. in context $\psi$) then *replace s by s'*. Formally, if we apply the rule $\psi \vdash s \leadsto s'$ to a family:

   ...#if ($\phi_1$) ...#if ($\phi_n$) ...; $s$; ... #endif... #endif...

where $\phi_1 \land \ldots \land \phi_n \implies \psi$, then we obtain the transformed program:

   ...#if ($\phi_1$) ...#if ($\phi_n$) ...; $s'$; ... #endif... #endif...

We write $\textit{Rewrite}(\overline{s}, \psi \vdash s \leadsto s')$ for the final transformed program $\overline{s'}$ obtained by repeatedly applying the rule $\psi \vdash s \leadsto s'$ on $\overline{s}$ and its transformed versions until we reach a point where this rule can not be applied (a fixed point of the rule). Note that rules of the form: true $\vdash s \rightsquigarrow s'$, are the most general and can be applied to any statement $s$ no matter whether $s$ is a top-level statement not nested within some #if or $s$ is nested somewhere deep within #if-s. This is due to the fact that any "$s$" can be written as: "#if (true) $s$ #endif" in the earlier case when $s$ is a top-level statement, and $\phi \implies$ true for any $\phi \in FeatExp$ in the latter case when $s$ is nested within #if-s with presence condition $\phi$.

---

[1] Here $\models$ denotes the standard satisfaction relation of propositional logic.

We start with three rules for eliminating configurable variable declarations. They involve duplicating code and variable renaming. The most straightforward way to handle renaming of variables in different contexts is by adding an *environment* $\delta$ as a parameter to the statements being transformed. We define an environment $\delta : Var \times FeatExp \to Var$ as a function mapping a given pair of a variable and a feature expression to a variable name. We write $\delta^{fe}(x) \subseteq FeatExp$ for the set of all feature expressions $\phi$ such that $\delta(x, \phi)$ is defined, i.e. $\delta^{fe}(x) = \{\phi \in FeatExp \mid (x, \phi) \in dom(\delta)\}$. We write $(s, \delta)$ to denote the result of simultaneously substituting $\delta(x, \phi)$ for each occurrence of any variable x in $s$ in the context (presence condition) that implies $\phi$.

**Conditional variable declaration.** This rule transforms a local variable conditionally declared within a given context $\psi \in FeatExp$:

$$\psi \vdash (\text{\#if } (\phi) \text{ var x:=}n \text{ in \#endif} s, \delta) \rightsquigarrow \text{var x}_{new}\text{:=}n \text{ in } (s, \delta[(x, \phi) \mapsto x_{new}]) \quad (1)$$

where $x_{new}$ is a fresh variable name that does not occur as a free variable in $s$ and $range(\delta)$.

**Conditional variable use.** The second rule handles the case when a local variable is used within a context $\psi \in FeatExp$. There are three cases to consider here.

$$\psi \vdash (y\text{:=}e[x], \delta) \rightsquigarrow (y\text{:=}e[\delta(x, \phi)], \delta) \quad (2.1)$$

if there exists an unique $\phi \in \delta^{fe}(x)$, such that $\psi \models \phi$. Here $e[x]$ means that the variable x occurs free in the expression $e$. The second case is when there are several $\phi_1, \ldots \phi_n \in \delta^{fe}(x)$, such that $sat(\phi_1 \wedge \psi), \ldots, sat(\phi_n \wedge \psi)$:

$$\psi \vdash (y\text{:=}e[x], \delta) \rightsquigarrow (\text{\#if } (\phi_1) \text{ y:=}e[\delta(x, \phi_1)] \text{ \#endif}; \ldots \text{\#if } (\phi_n) \text{ y:=}e[\delta(x, \phi_n)] \text{ \#endif}, \delta)$$
$$(2.2)$$

Otherwise, meaning that for all $\phi \in \delta^{fe}(x)$ it follows that $unsat(\phi \wedge \psi)$, we have:

$$\psi \vdash (y\text{:=}e[x], \delta) \rightsquigarrow (y\text{:=}e[x], \delta) \quad (2.3)$$

**Conditional variable define.** The third rule applies when a local variable is assigned to within a context $\psi \in FeatExp$. There are three cases to consider here as well.

$$\psi \vdash (x\text{:=}e, \delta) \rightsquigarrow (\delta(x, \phi)\text{:=}e), \delta \quad (3.1)$$

when there exists an unique $\phi \in \delta^{fe}(x)$, such that $\psi \models \phi$.

$$\psi \vdash (x\text{:=}e, \delta) \rightsquigarrow (\text{\#if } (\phi_1) \delta(x, \phi_1)\text{:=}e \text{ \#endif}; \ldots \text{\#if } (\phi_n) \delta(x, \phi_n)\text{:=}e \text{ \#endif}), \delta \quad (3.2)$$

when there are $\phi_1, \ldots \phi_n \in \delta^{\text{fe}}(x)$, such that $\text{sat}(\phi_1 \wedge \psi), \ldots, \text{sat}(\phi_n \wedge \psi)$. Otherwise,

$$\psi \vdash (\text{x:=}e, \delta) \rightsquigarrow (\text{x:=}e, \delta) \tag{3.3}$$

After applying the above three rules, all local variable declarations that are conditionally defined (#if ($\phi$) var x:=n in #endif$s$) are resolved. The transformed program contains only #if-s where statements are conditionally defined.

**Conditional statement elimination.** The set of valid configurations $\mathbb{K}$ can be equated to a propositional formula [4], say $\kappa \in \textit{FeatExp}$, such that $\kappa = \vee_{k \in \mathbb{K}} k$. The last rule simply replaces #if-s with ordinary if-s whose guards are strengthen with the feature model $\kappa$, thus taking into account only valid configurations $\mathbb{K}$ of a family.

$$\psi \vdash \#\text{if } (\phi) \ s \ \#\text{endif} \rightsquigarrow \text{if } (\phi \wedge \kappa) \text{ then } s \text{ else skip} \tag{4}$$

Note that we omit to write the environment $\delta$ in rules that do not use it explicitly (e.g. rules (4), (5)). Let $\delta_0 = [\,]$ be the empty environment. Let $\textit{Rewrite}^{\text{preserve}}(\text{pre-t}(\bar{s}), \delta_0)$ be the final transformed program $\bar{s'}$ obtained from the pre-transformed program pre-t($\bar{s}$) by applying the rules (1)–(3), and then the rule (4). The following result shows that the set of final answers from terminating computations of $\bar{s'}$ coincides with the union of final answers from terminating computations of all variants from $\bar{s}$.

**Theorem 1.** *Let $\bar{s'} = \textit{Rewrite}^{\text{preserve}}(\textit{pre-t}(\bar{s}), \delta_0)$. We have: $[\![\bar{s'}]\!] = \bigcup_{k \in \mathbb{K}} [\![\pi_k(\bar{s})]\!]$.*

*Proof.* First, we show that $\textit{Rewrite}^{\text{preserve}}(\text{pre-t}(\bar{s}), \delta_0)$ terminates. This is due to the fact the number of if-s in pre-t($\bar{s}$ is finite, and by iteratively applying rules (1)–(3) we eliminate all #if ($\phi$) var x:=n in #endif$s$; whereas by applying rule (4) afterwards we eliminate all #if ($\phi$) $s$ #endif. Subsequently, for each rule (1)–(3) and (4), the above result can be proved by structural induction. □

We now present an optimization rule, which is applied before the rules (1)–(4) for eliminating if-s. The correctness of our transformation does not depend on it, but we can use it for achieving faster convergence and smaller transformed programs. In our implementation, we use many such optimization rules.

**Guard inlining.** This rule collapses two sequentially composed #if-s with mutually exclusive presence conditions $\phi_0$ and $\phi_1$ (i.e. $\phi_0 \wedge \phi_1 \equiv \text{false}$) that conditionally enable the same statement $s$ into one #if that conditionally enables $s$:

$$\psi \vdash \#\text{if } (\phi_0) \ s \ \#\text{endif}; \#\text{if } (\phi_1) \ s \ \#\text{endif} \rightsquigarrow \#\text{if } (\phi_0 \vee \phi_1) \ s \ \#\text{endif} \tag{5}$$

**Example 2.** *We present the transformation rules on a program family with $\mathbb{F} = \{A, B\}$ and $\mathbb{K} = \{A \wedge B, A \wedge \neg B, \neg A \wedge B, \neg A \wedge \neg B\}$.*

$\big(\#\text{if } (A) \text{ var x:=2 in } \#\text{endif}\#\text{if } (\neg A) \text{ var x:=5 in } \#\text{endif}\#\text{if } (B) \text{ y:=x } \#\text{endif}, \delta_0\big)$
$\overset{(1)}{\rightsquigarrow} \text{var } x_1\text{:=2 in } \big(\#\text{if } (\neg A) \text{ var x:=5 in } \#\text{endif}\#\text{if } (B) \text{ y:=x } \#\text{endif}, [(x, A) \mapsto x_1]\big)$
$\overset{(1)}{\rightsquigarrow} \text{var } x_1\text{:=2 in var } x_2\text{:=5 in } \big(\#\text{if } (B) \text{ y:=x } \#\text{endif}, [(x, A) \mapsto x_1, (x, \neg A) \mapsto x_2]\big)$
$\overset{(2.2)}{\rightsquigarrow} \text{var } x_1\text{:=2 in var } x_2\text{:=5 in } \#\text{if } (B) \ \#\text{if } (A) \text{ y:=}x_1; \#\text{endif}\#\text{if } (\neg A) \text{ y:=}x_2 \ \#\text{endif } \#\text{endif}$
$\overset{(4)}{\rightsquigarrow} \text{var } x_1\text{:=2 in var } x_2\text{:=5 in } \textit{if}(B) \textit{ then if } (A) \textit{ then } y\text{:=}x_1 \textit{ else skip};$
$\qquad\qquad\qquad\qquad\qquad\qquad\quad \textit{if } (\neg A) \textit{ then } y\text{:=}x_2 \textit{ else skip}; \textit{else skip}$

## 5 Implementation

We have developed a tool, called C Reconfigurator, which implements variability-related transformations for the C language. All transformations are implemented using Xtend [2]. The C Reconfigurator tool is available from: https://github.com/models-team/c-reconfigurator. It calls variability-aware parser SuperC [20] to parse code with preprocessor annotations, which uses Binary Decision Diagrams (BDD's) for encoding feature expressions and for decisions during the parsing process. SuperC returns an AST with variability, in which variability is reflected with choice nodes over feature expressions. In particular, a choice node is a node with two children, such that the left child of the choice node is included in the result of those configurations for which the given feature expression is satisfied; otherwise the right child of the choice node is included in the parsing result when the feature expression is not satisfied. We apply our variability-related transformation rules as described in Section 4 on AST's with variability obtaining an ordinary AST, which is subsequently translated into a single C program. Since IMP is a subset of C, all rewritings described in Section 4 transfer directly to C. We now discuss how a selection of other interesting C constructs, which are not present in IMP, are handled by our tool.

Variables declared with optional types are very common in C. For example, we have x-bit integers on x-bit machines. We handle them in a similar way as configurable variable declarations in rules (1)–(3). First, we rename and duplicate the variable declaration, then at each point where the variable is used we transform the code such that the used variable refers to the correct configuration name. For example,

$\quad$ #if $(A)$ int #else float #endif x=0;
$\quad$ x = x+1;

will be transformed into:

$\quad$ int $x_1 = 0$; float $x_2 = 0$;
$\quad$ #if $(A)$ $x_1 = x_1$+1; #else $x_2 = x_2$+1; #endif

Note that if optional local variables are initialized by non-constant expressions, then we split their transformation into two parts: declaration which is performed by renaming and duplication, followed by initialization where all optional variables refer to the correct configuration.

Optional (configurable defined) functions are important since all statements in C are inside some function. If conditionally defined code occurs in the function body, then it will be transformed using the corresponding rules. For example,

$\quad$ int $f$ (int x) {return #if $(A)$ x++ #else 0 #endif; }

will be transformed into:

$\quad$ int $f$ (int x) {return $A$ ? x++ : 0; }

---

[2] http://www.eclipse.org/xtend/.

If the function signature is configurable, then we use renaming plus duplication as in rules (1)–(3) for handling configurable variable declarations. For example, the code:

    int $f$(#if($A$) int #else float #endif x) {...}
    ... $f$(5)...

will be transformed into:

    int $f_1$(int x) {...}
    int $f_2$(float x) {...}
    ... #if($A$) $f_1$(5) #else $f_2$(5) #endif ...

Arrays with optional size are also possible in real-world C programs. They usually emerge via constant macros with conditional definitions. For example, the code

    int a[#if($A$) 10 #else 15 #endif ];
    a[5] = 0;

will be transformed into:

    int $a_1$[10]; int $a_2$[15];
    #if($A$) $a_1$[5] = 0; #else $a_2$[5] = 0; #endif

All other variability patterns that we met in our examples, such as configurable fields in struct-s and pointers, are also handled similarly: first by using renaming and duplication, then by modifying all references to the given pattern such that the use always refers to the correct definition. Consider the following code with pointers:

    int a = 10; int ∗ p = &a; #if($A$) p = $null$; #endif (∗p)++

will be transformed into:

    int a = 10; int ∗ p = &a; if($A$) p = $null$; (∗p)++

Hence, we obtain a variability bug whenever the feature $A$ is enabled.

**Remark.** We can see that most of the variability patterns are handled using renaming plus duplication. In the worst case, this may cause exponential growth of the transformed program in the number of used features. However, in practice this does not happen often (see Table 3 for some data from real files). Namely, variability patterns usually depend on a few features, so only a few new definitions are used. Also we apply several optimization rules, which eliminate all definitions that do not correspond to a valid configuration. Finally, the evaluation results in Section 6 show that the analysis time for such transformed programs is comparable to single programs. This is due to the fact that transformed programs are not increased significantly and the analysis tools we use (FRAMA-C, CLANG, LLBMC) are very optimized and mature.

## 6 Evaluation

We evaluate our reconfiguration technique based on variability transformations and single-program verification oracles on several real-world C case studies. The evaluation aims to show that we can use state-of-the-art single-program verification tools to verify realistic C program families using variability-related transformations. To do so, we ask the following research questions:

- How precise is our technique? **(RQ1)**
- How efficient is the verification oracle to identify variability bugs after transforming the code using our technique? **(RQ2)**

In particular, we want to reproduce the variability bugs reported in [1, 28] using various verification oracles applied on transformed programs, which are obtained using our tool. We use FRAMA-C [27], CLANG [9] and LLBMC [30] as our verification oracles. FRAMA-C is a framework for modular static (dataflow) analysis of C programs. The CLANG project includes the Clang compiler front-end and the Clang static analyzer for several programming languages, including C. LLBMC (the low-level bounded model checker) is a software model checking tool for finding bugs in C programs.

### 6.1 Subject Files and Experimental Setup

All transformations are applied using the C RECONFIGURATOR tool as described in Section 5. We investigate precision and performance in finding real variability bugs extracted from three benchmarks: Linux, BusyBox and Libssh. In particular, we use simplified bugs from the VDBb [3] database that are found in the Linux kernel files [1] and in BusyBox. Abal et al. [1] created a simplified version of a program for each bug they found by capturing the same essential behavior (and the same problem) as in the original bug. Simplified bugs are independent of the kernel code and the corresponding programs were derived systematically from the error trace. In addition, we use real variability bugs from Libssh provided by Medeiros et al. [28].

Table 1 presents the characteristics of the subject files we analyzed in our empirical study. We list: the file id, bug type, number of features ($|\mathbb{F}|$), number of valid configurations ($|\mathbb{K}|$), lines of code, the size in KB of the files before (with #ifdef-s) and after (without #ifdef-s) our transformations, and commit hash (clickable) for each project. This collection consists of a diverse set of bug types such as null pointer dereferences, buffer overflow, and uninitialized variable. In total, we have 11 distinct kinds of bugs. The number of features per file varies from one to seven. In addition, the number of lines of code ranges from 12 to 165 for the simplified files (from VBDb), and from 1404 to 2959 for real files (from Libssh). After the transformation, the biggest increase in size of almost 8 times can be observed for FILE ID 7. This is due to the fact that this file has seven different features and several variability patterns that depend on them. In most of the other cases the size increase is not very big.

---

[3] http://VBDb.itu.dk.

# Effective Analysis of C Programs by Rewriting Variability

| File id | Bug type | $|\mathbb{F}|$ | $|\mathbb{K}|$ | LOC | Size KB before | Size KB after | Hash |
|---|---|---|---|---|---|---|---|
| VBDb Linux files ||||||||
| 1 | null pointer deref. | 5 | 24 | 165 | 2.9 | 4.3 | 76baeeb |
| 2 | null pointer deref. | 3 | 6 | 112 | 1.9 | 2.5 | f7ab9b4 |
| 3 | null pointer deref. | 4 | 8 | 55 | 0.9 | 1.0 | ee3f34e |
| 4 | null pointer deref. | 3 | 6 | 34 | 0.5 | 0.6 | 6252547 |
| 5 | buffer overflow | 1 | 2 | 58 | 1.0 | 1.2 | 8c82962 |
| 6 | buffer overflow | 1 | 2 | 33 | 0.6 | 0.7 | 60e233a |
| 7 | read out of bounds | 7 | 63 | 69 | 1.1 | 8.4 | 0f8f809 |
| 8 | uninitialized var. | 2 | 4 | 54 | 0.8 | 1.0 | 7acf6cd |
| 9 | uninitialized var. | 1 | 2 | 54 | 1.0 | 1.1 | bc8cec0 |
| 10 | uninitialized var. | 1 | 2 | 53 | 0.8 | 1.0 | 30e0532 |
| 11 | uninitialized var. | 2 | 4 | 38 | 0.9 | 1.2 | 1c17e4d |
| 12 | uninitialized var. | 2 | 4 | 26 | 0.3 | 0.5 | e39363a |
| 13 | undefined symbol | 4 | 14 | 25 | 0.4 | 0.6 | 7c6048b |
| 14 | undefined symbol | 2 | 4 | 20 | 0.3 | 0.5 | 2f02c15 |
| 15 | undefined symbol | 2 | 4 | 20 | 0.3 | 0.5 | 6515e48 |
| 16 | undefined symbol | 2 | 4 | 19 | 0.3 | 0.5 | 242f1a3 |
| 17 | undeclared identifier | 3 | 8 | 37 | 0.6 | 1.0 | 6651791 |
| 18 | undeclared identifier | 2 | 4 | 20 | 0.3 | 0.4 | f48ec1d |
| 19 | wrong # of args | 1 | 2 | 12 | 0.2 | 0.4 | e67bc51 |
| 20 | multiple funct. defs | 2 | 4 | 21 | 0.3 | 0.8 | e68bb91 |
| 21 | dead code | 1 | 2 | 19 | 0.2 | 0.3 | 809e660 |
| 22 | incompatible type | 2 | 4 | 27 | 0.4 | 0.7 | d6c7e11 |
| 23 | assertion violation | 2 | 4 | 79 | 1.5 | 1.8 | 63878ac |
| 24 | assertion violation | 2 | 4 | 75 | 1.1 | 1.2 | 657e964 |
| 25 | assertion violation | 2 | 4 | 41 | 0.6 | 0.7 | 0988c4c |
| VBDb BusyBox files ||||||||
| 26 | null pointer deref. | 1 | 2 | 28 | 0.4 | 0.7 | 199501f |
| 27 | null pointer deref. | 2 | 4 | 24 | 0.4 | 0.6 | 1b487ea |
| 28 | uninitialized var. | 2 | 4 | 28 | 0.4 | 0.7 | b273d66 |
| 29 | undefined symbol | 1 | 2 | 42 | 0.8 | 0.9 | cf1f2ac |
| 30 | undefined symbol | 2 | 4 | 27 | 0.4 | 0.6 | ebee301 |
| 31 | undeclared identifier | 1 | 2 | 35 | 0.5 | 0.8 | 5275b1e |
| 32 | undeclared identifier | 1 | 2 | 19 | 0.3 | 0.4 | b7ebc61 |
| 33 | incompatible type | 3 | 8 | 46 | 0.9 | 1.5 | 5cd6461 |
| Real Libssh files ||||||||
| 34 | null pointer deref. | 6 | 48 | 1404 | 34.8 | 32.6 | 0a4ea19 |
| 35 | null pointer deref. | 4 | 4 | 1428 | 44.1 | 31.9 | fadbe80 |
| 36 | uninitialized var. | 3 | 4 | 2959 | 72.4 | 77.6 | 2a10019 |

■ **Table 1** Characteristics of the benchmark files.

A. F. Iosif-Lazar, J. Melo, A. S. Dimovski, C. Brabrand, A. Wasowski| | Frama-C | | | |
|---|---|---|---|---|
| ID | BUGGY VARIANT | | RECONFIGURED | ALL |
| | y/n | time | y/n  time | time |
| VBDb Linux files | | | | |
| 1 | ✓ | 218 | ✓  235 | 5602 |
| 2 | ✓ | 220 | ✓  225 | 1394 |
| 3 | ✓ | 215 | ✗  236 | 1918 |
| 4 | ✓ | 218 | ✓  224 | 1379 |
| 5 | ✓ | 218 | ✓  227 | 488 |
| 6 | ✓ | 213 | ✓  227 | 463 |
| 7 | ✓ | 218 | ✓  225 | 14381 |
| 8 | ✓ | 241 | ✓  250 | 918 |
| 9 | ✓ | 224 | ✓  230 | 462 |
| 10 | ✓ | 216 | inc  224 | 460 |
| 11 | ✓ | 234 | ✓  224 | 917 |
| 12 | ✓ | 216 | inc  227 | 914 |
| 13 | ✓ | 239 | ✓  248 | 3194 |
| 14 | ✓ | 237 | ✓  244 | 905 |
| 15 | ✓ | 224 | ✓  248 | 906 |
| 16 | ✓ | 213 | ✓  222 | 910 |
| 17 | ✓ | 216 | ✓  230 | 3823 |
| 18 | ✓ | 210 | ✓  224 | 901 |
| 19 | ✓ | 210 | ✓  224 | 452 |
| 20 | ✓ | 213 | ✗  228 | 907 |
| 21 | ✓ | 239 | ✗  240 | 458 |
| VBDb BusyBox files | | | | |
| 26 | ✓ | 230 | ✓  234 | 484 |
| 27 | ✓ | 224 | ✓  234 | 959 |
| 28 | ✓ | 237 | inc  237 | 957 |
| 29 | ✓ | 230 | ✓  236 | 481 |
| 30 | ✓ | 231 | ✓  228 | 968 |
| 31 | ✓ | 220 | ✓  228 | 486 |
| 32 | ✓ | 216 | ✓  224 | 477 |

**(a)** VBDb files using Frama-C.

| | Clang/LLBMC | | | |
|---|---|---|---|---|
| ID | BUGGY VARIANT | | RECONFIGURED | ALL |
| | yes/no | time | yes/no  time | time |
| VBDb Linux files | | | | |
| 22 | ✓ | 21 | ✓  23 | 91 |
| 23 | ✓ | 4 | ✓  10 | 10 |
| 24 | ✓ | 3 | ✓  7 | 11 |
| 25 | ✓ | 3 | ✓  5 | 8 |
| VBDb BusyBox files | | | | |
| 33 | ✓ | 27 | ✓  31 | 222 |

**(b)** VBDb files using Clang (files 22 and 33) and LLBMC (files 23, 24, and 25).

| | Clang/LLBMC | | | |
|---|---|---|---|---|
| ID | BUGGY VARIANT | | RECONFIGURED | ALL |
| | yes/no | time | yes/no  time | time |
| 34 | ✓ | 1526 | ✓  1702 | 17029 |
| 35 | ✓ | 1591 | ✓  1804 | 5917 |
| 36 | ✓ | 112 | ✓  144 | 448 |

**(c)** Libssh files using Clang (file 36) and LLBMC (files 34 and 35).

■ **Table 2** Verification results for the benchmark files. Times in milliseconds (ms).

All experiments were executed on a Kubuntu VM (64bit, 4 CPUs), Intel®Core$^{TM}$ i7-3720QM CPU running at 2.6GHz with 12GB RAM memory. The performance numbers reported constitute the median runtime of fifty independent executions.

### 6.2 Results

We now present the results of our empirical study and discuss the implications. All experiment materials are available online at https://github.com/models-team/c-reconfigurator-test. Before we proceed, we stress that we only evaluate bugs that are detectable by the verification tools on the erroneous variant code.

1-15

**Effective Analysis of C Programs by Rewriting Variability**

**Simplified files.**  Table 2a shows the results of verifying our benchmark files which contain known bugs by using Frama-C. The table has three main columns: BUGGY VARIANT, RECONFIGURED, and ALL that depict the tool results on the buggy variant code, on the reconfigured program family code, and on all valid variants from $\mathbb{K}$ analyzed one by one (in a brute force fashion), respectively. Each checkmark ($\checkmark$) means that the same bug was found in both the buggy variant and reconfigured program by the verification tool. Otherwise, the result is either *x*—bug not found in the reconfigured program, or *inc*—inconclusive which means that Frama-C was able to detect a bug in the reconfigured program that is different from the bug in the product variant. In the case of brute force approach (ALL), we consider the analyses times of all valid variants regardless of whether they contain a bug or not.

In terms of precision, our C Reconfigurator tool transforms the family code by preserving the erroneous traces from the buggy variant in most cases. For instance, Frama-C could detect 22 (78%) bugs from the simplified benchmark files (28 in total) after reconfiguring the files using our tool. Besides that, the C Reconfigurator preserves a variety of bug types such as buffer overflow and uninitialized variable. Still, for different types of bugs the success rate depends on the tool which may or may not detect them. For example, our technique is able to transform a file containing a memory leak error, but Frama-C does not have any analysis to identify it.

In three specific cases (cf. FILE IDS 10, 12 and 28), Frama-C did not report the original bug as an error, but it did detect that some variable might be uninitialized in some conditions. This happens because Frama-C performs a *may* value analysis for finding uninitialized variables. A *may* analysis describes information that may possibly be true along one path to the given program point and, thus in our case, computes a superset of all uninitialized variables in all variants. So the reported variable may not match with the one in the buggy variant. We marked these three cases as *inc*—inconclusive in the table. Still the verification oracle reports that there might be an error in the reconfigured code.

In addition, the verification tool could not identify the required bug in the reconfigured file in three cases (cf. FILE IDS 3, 20 and 21). For example, file 21 contains dead code, which is a function (`do_sect_fault()`) that is never called when feature ARM is enabled (see the code snippet in Fig. 4, left column). The C Reconfigurator transforms the code by changing the `#ifdef` into ordinary `if` condition, making the function available for the transformed single program (i.e., the function is not dead any more), as shown in the code snippet in Fig. 4 (right column). The other two cases are similar to this one in the sense that the C Reconfigurator makes feature code explicit to the entire program family.

Generally speaking, if one variant does not use a variable/function, but another does, then the reconfigured code will use the variable/function and the error will be hidden (like in the example above). This happens due to the limitations of variability encoding, especially because we cannot preprocess the reconfigured code to filter out the irrelevant features for a particular variant. In a reconfigured code, all variants are encoded as a single program (see Section 6.4 for more discussion).

We now consider the remaining simplified files. We use Clang and LLBMC to analyze only the other types of bugs (incompatible type and assertion violation) that

```
int do_sect_fault(){        int do_sect_fault(){
   return 0;                   return 0;
}                           }
int main(){
   #ifndef ARM             int main(){
      do_sect_fault();        if (! ARM)
   #endif                        do_sect_fault();
   return 0;                  return 0;
}                           }
```

■ **Figure 4** File 21 - Before (left) and after (right) our transformations

Frama-C cannot handle. We treat Clang/LLBMC as one verification oracle, since we first need to compile and emit llvm code with Clang in order to analyze it using LLBMC. So, we do not make difference in reporting whether the bug was found by Clang during the compilation or afterwards by LLBMC.

Table 2b, similarly to Table 2a, shows the results of verifying both the buggy variant and the reconfigured code using Clang and LLBMC. We also report the analysis time of the brute force approach in the column ALL. As we can see, all bugs were found by Clang/LLBMC in the reconfigured version. We can thus confirm that our C Reconfigurator tool transforms the family code by preserving the erroneous traces from the buggy variant. We are now ready to answer RQ1 on the precision of our technique. Based on analyzing 33 simplified variability bugs from Linux and BusyBox, we find that:

> Answer RQ1 (precision). The C Reconfigurator enables single-program verification tools such as Frama-C, Clang, and LLBMC to **successfully** detect *most* of the simplified variability bugs on the reconfigured code, obtained from the Linux and BusyBox benchmark files.

We now turn to evidence regarding research question RQ2 (performance). We evaluate performance of the verification tools to identify the given variability bugs. Tables 2a and 2b show time needed for the verification tools to analyze the buggy variant code (BUGGY VARIANT column) and the reconfigured program family code (RECONFIGURED column). We can see that the analysis times in both cases are similar although reconfigured code is bigger in size. In fact, Frama-C takes less than half a second to analyze each file regardless whether it is a variant or a reconfigured file. For instance, Frama-C analyzes file 1 in 218 and 235 milliseconds on the variant code and on the reconfigured program family code, respectively. Recall that file 1 contains a null pointer dereference and has five features. If we apply the brute force approach (ALL column), which analyzes all variants individually one by one, to this file using Frama-C it takes 5,602 ms, since the number of configurations is 24. In this way, we obtain significant speed-up to verify the program family using our approach. We also obtain similar results in terms of performance using Clang/LLBMC (see Tables 2b and 2c). In general, the performance of analyzing a reconfigured code is similar to analyzing only one variant, which gives us a speed-up proportional to the number of valid variants of a program family. Overall, we answer the second research question (RQ2) by observing that:

> Answer RQ2 (performance). The C Reconfigurator speeds-up the family-based analysis via single-program verification tools, so that we can **efficiently** detect simplified variability bugs on the reconfigured code, obtained from the VBDb benchmark.

**Real files.** We now consider real files to confirm our previous observations with respect to precision and performance. Table 2c presents the results of analyzing three real files from the Libssh project using Clang and LLBMC.[4] These files contain two types of bugs: null pointer dereference and uninitialized variable. Each file has at least three distinct features.

We can see that our C Reconfigurator transforms the family code by preserving the erroneous traces from the buggy variant even for complex and large files. In fact, the verification tool (Clang/LLBMC) found the same bug (from the buggy variant code) on the reconfigured code in all three cases. From this preliminary evidence, we thus confirm that our technique enables single-program verification oracles to successfully detect variability bugs on the reconfigured code, obtained from complex and real files.

Regarding performance, we can still see the similarity in verifying a variant code and a reconfigured one. For example, Clang/LLBMC took 1,5 sec to analyze file 34 in the single variant version, whereas in the reconfigured version, the tool analyzed it in 1,7 sec. We can also observe a speed-up of the family-based analysis using the C Reconfigurator and single-program verification tools by a factor of the number of valid variants compared to the brute force approach. We conclude that:

> Summary. All single-program verification tools (Frama-C, Clang, LLBMC) detect **successfully** and **efficiently** most of the variability bugs on the reconfigured code as well as on the single variant code.

### 6.3 Threats to Validity

**Internal validity.** Verifying semantics preservation in a complex transformation is a very hard problem [22, 2]. We manually verified the correctness of the C Reconfigurator on the simplified VBDb files by comparing the original and the reconfigured files side-by-side, which leaves space for human error. For the larger real files we were not able to determine if the C Reconfigurator preserved semantics for all variants on the entire file due to the complex configuration space, but instead we focused on the functions involved in producing/reproducing the bug. We mitigate this threat by relying on the results of our evaluation which show the effectiveness of conventional single-program analysis tools to identify the same bugs in the reconfigured code version as in the buggy single varaints.

---

[4] We do not report results from Frama-C on the real files because Frama-C could not handle them.

**External validity.** From our preliminary evaluation, we show that our technique transforms the program family code by preserving the erroneous traces from the buggy variant. However, we acknowledge that our transformations were not tested under the entirety of the C language, but only on the subset used in the VBDb and Libssh files presented here. The C Reconfigurator though can be extended with extra rules to deal with other cases that we did not face in our benchmark files. Worst case exponential growth of transformed programs can happen, even though we have not observed it in our subject files.

## 6.4 Discussion

The main limitation of our transformation based approach is that we may not obtain conclusive results for all individual variants, thus losing some precision. This is due to the fact that our transformed program contains all possible paths that may occur in any variant. However, the precision loss depends on the particular analysis we use.

Consider the case of model checking. Since (single-system) model checkers stop once a single counter-example is found in the model, we can use our approach to find a variability bug which occurs in some subset of valid variants but we will not be able to report conclusive results (whether the given property is satisfied or not) for the rest of the valid variants. To overcome this issue, we may repeat our technique on the remaining variants for which no conclusive results were reported in the previous iteration.

Consider the case of *must* dataflow analysis (e.g., available expressions, very busy expressions). In this case, the result in a given program point contains only the common results found on all execution paths to that point. Thus, the analysis result for the transformed program will contain only the results that occur in all variants. For example, for available expressions analysis we may obtain less available expressions than there are in any single variant. The available expression analysis determines which expressions must have already been computed, and not later modified, on all paths to a program point [32]. This information can be used to avoid re-computation of an expression. Consider the program family:

$$\textsf{x} := a + b; \textsf{while } (\textsf{y} > a + b) \textsf{ do } \{\, \textsf{\#ifdef } (A) \textsf{ y} := \textsf{y} - 1 \textsf{ \#else } a := a + 1 \textsf{ \#endif}\,\}$$

The expression $a + b$ is available at the guard of the while loop for variants satisfying $A$, so it needs not be re-computed for them. However, in the transformed program we have paths from all variants, so the expression $a + b$ is modified by the assignment $a := a + 1$ in a path coming from variants $\neg A$. Therefore, the analyzer will not report this expression as available at the guard of the loop for the transformed program.

Consider the case of *may* dataflow analysis (e.g., reaching definitions, live variables, uninitialized variables). In this case, the result in a given program point contains the results found on at least one execution path to that point. Thus, the analysis result for the transformed program will contain all results that occur in at least one variant. For example, for live variables analysis, we may obtain more live variables than there are in any single variant. The live variables analysis determines which variables may be live at a program point, that is there is a path from the program point to a use of the variable that does not redefine it [32]. This information can be used as a basis for dead code elimination. If a variable is not live at the exit from an assignment to the variable, then that assignment can be eliminated. Consider the program family:

$$x := 5; y := 1; \#\text{ifdef}\ (A)\ x := 1\ \#\text{else}\ x := x + 1\ \#\text{endif}$$

The variable x is not live at the exit from the first assignment x := 5 for variants satisfying $A$. Therefore, the assignment x := 5 is redundant for those variants. However, x is live for $\neg A$ variants, so it will be live after the first assignment for the transformed program as well. Thus, we cannot eliminate this assignment in the transformed program. This is also the reason why Frama-C does not identify the variability bug for files 3, 20 and 21.

## 7 Related work

Recently, formal analysis and verification of program families have been a topic of considerable research. The challenge is to develop efficient techniques that work at the level of program families, rather than the level of single programs. There are two main approaches to address this issue: (1) to develop dedicated variability-aware (family-based) techniques and tools; (2) to use specific simulators and encodings which transform program families into single programs that can be analyzed by the standard single-program verification tools. The two approaches have different strengths and weaknesses. The advantage of (1) is that precise (conclusive) results are reported for every variant, but the disadvantage is that their implementation can be tedious and labor intensive. On the other hand, the approaches based on (2) re-use existing tools from single-program world, but some precision may be lost when interpreting the obtained results.

**Specifically designed variability-aware techniques.** Various lifted techniques have been proposed which lift existing single-program verification techniques to work on the level of program families. This includes lifted syntax checking [25, 20], lifted type checking [24, 8], lifted static analysis [7, 6, 31], lifted model checking [10, 14], etc. TypeChef [25] and SuperC [20] are variability-aware parsers, which can parse languages with preprocessor annotations. The results are ASTs with variability nodes. The difference between these two approaches is that feature expressions are represented as formulae in TypeChef, and as BDD's in SuperC. TypeChef has also implemented some variability-aware dataflow analyses. Several approaches have been proposed for type checking program families directly. In particular, lifted type checking for Featherweight Java was presented in [24], whereas variational lambda calculus was studied in [8]. Lifted model checking for verifying variability intensive systems has been introduced in [10]. SNIP, a specifically designed family-based model checker, is implemented for efficient verification of temporal properties of such systems. The input language to SNIP is fPromela, which represents a variability-aware extension of the known Promela language for the (single-system) SPIN model checker [21]. fPromela uses an #ifdef-like statement for encoding multiple variants, which rep-resents a nondeterministic "if" statement guarded by features expressions that are used to specify what system parts are included (resp., excluded) for which variants. An approach for lifted software model checking using game semantics has been introduced in [14]. It verifies safety of #ifdef-based second-order program families containing undefined components, which are compactly represented using symbolic game semantics models [13, 12]. Brabrand et al. [7] and Midtgaard et al. [31] show how to lift any single-program dataflow analysis from the monotone framework to work on the level of program families. The obtained lifted dataflow analyses are much faster than ones based on the naive variant-by-variant approach that generates and analyzes all variants one by one. Another efficient implementation of lifted analysis formulated within the IFDS framework for inter-procedural distributive environments has been proposed in SPL$^{\text{LIFT}}$ [6]. In order to speed-up the lifted verification techniques, variability abstractions have been introduced in [17, 18, 15, 16]. They tame the exponential blowup caused by the large number of features and variants in a program family. In this way, variability abstractions enable deliberate trading of precision for speed in the context of lifted (monotone) data-flow analysis [17, 18] and lifted model checking [15, 16].

**Lifting by simulation.** Variability encoding [37] and configuration lifting [33] are based on generating a product-line *simulator* which simulates the behaviour of all products in the product line. Then, an existing off-the-shelf single-program analyzer is used to verify the generated product-line simulator, which represents a single program. The work in [37] defines variability encoding on the top of TypeChef parser for C and Java program families. They have applied the results of variability encoding to testing [26], model checking [3], and deductive verification [36]. Compared to [37], our approach has the following distinguished characteristics. C Reconfigurator is aimed at transforming C program families and uses SuperC as a back-end tool. We show transformation rules and their correctness with respect to a minimal C-like imperative (state-based) language, whereas in [37] the rules and their correctness is shown with respect to Featherweight Java. C is a language much wider used in industry for variability than (Featherweight)Java. Also, we do not have to rely on object-oriented encodings to make the variability-transformations work. We evaluate our approach with several state-of-the-art single-program verification tools for finding real variability bugs on real-world C programs (both on large and sanitized files). The academic examples (e-mail, elevator, mine-pump) considered by Apel et al. [3] are considerably smaller than those presented here; and they are focussed on verifying specific class of bugs: undesired feature interactions (using CPAchecker [5]), whereas we consider here various types of more severe bugs that occur in practice. In this way, the external validity of our experiments is considerably broader. Yet another difference is that the work in [3] considers product lines implemented using compositional approaches, where all features are modeled as separate and composable units. In contrast, we consider here annotative product lines based on #ifdef-s, which is a common way of implementing variability in industry.

## 8 Conclusion

We have proposed variability-related transformations to translate program families into single programs without variability. The transformed programs can then be effectively analyzed using various single-program analyzers. The evaluation confirms that some interesting variability bugs can be found in real-world C programs in this way. As a future work, we plan to extend our evaluation and consider more verification oracles as well as different practical case studies. We derive several observations from the attempt to verify, analyze, and find bugs in realistic C programs. We hope that our technique will be useful for future builders of analysis tools.